# Agile Missile Controller Based on Adaptive Nonlinear Backstepping Control

Chang-Hun Lee[1], Tae-Hun Kim[2] and Min-Jea Tahk[3]
*Korea Advanced Institute of Science and Technology(KAIST), Daejeon, 305-701, Korea*

**This paper deals with a nonlinear adaptive autopilot design for agile missile systems. In advance of the autopilot design, an investigation of the agile turn maneuver, based on the trajectory optimization, is performed to determine state behaviors during the agile turn phase. This investigation shows that there exist highly nonlinear, rapidly changing dynamics and aerodynamic uncertainties. To handle of these difficulties, we propose a longitudinal autopilot for angle-of-attack tracking based on backstepping control methodology in conjunction with the time-delay adaptation scheme.**

## Notice

---

[1] Ph. D, Department of Aerospace Engineering, Korea Advanced Institute of Science and Technology(KAIST), Kuseong Yuseong, Daejeon, 305-701, Korea/chlee@fdcl.kaist.ac.kr.
[2] Ph. D, Department of Aerospace Engineering, Korea Advanced Institute of Science and Technology(KAIST), Kuseong Yuseong, Daejeon, 305-701, Korea/thkim@fdcl.kaist.ac.kr.
[3] Professor, Department of Aerospace Engineering, Korea Advanced Institute of Science and Technology(KAIST), Kuseong Yuseong, Daejeon, 305-701, Korea/mjtahk@fdcl.kaist.ac.kr.

# I. Introduction

The demand for highly maneuverable missile systems has grown recently because of their usefulness in air-to-air combat scenarios. These missile systems, called agile missiles, are generally operating in the high angle-of-attack mode to achieve agile maneuverability. The flight envelope of the agile missile can be generally classified into three phases [1, 2]: launch and separation from the aircraft, the near-180 degree heading reversal maneuver (agile turn), and terminal homing. For each flight phase, suitable autopilot design strategies and techniques are necessary. The autopilot design problems for the separation and the terminal homing phases have been extensively studied and are well-understood via the autopilot design of conventional missile systems. However, the autopilot design for the agile turn phase is relatively less well-understood, and there exist crucial difficulties: the handling of highly nonlinear, rapidly changing missile dynamics and aerodynamics parameter uncertainties due to the poor measurement of aerodynamic data and the un-quantified coupling effect of aerodynamics in the high angle-of-attack. Over the course of several years, there have been various approaches to the autopilot design of the agile turn in order to solve these difficulties, ranging from linear control to nonlinear control methodologies.

In this paper, we deal with the angle-of-attack controller design for the agile turn phase in the pitch plane, using nonlinear backstepping control [14, 15] and time-delay control [16] methodologies. The backstepping control has been successfully applied to the flight control system from previous studies [11, 12, 17]. Accordingly, the backstepping control methodology is adopted for the main structure of the proposed autopilot. In the proposed method, the time-delay control is used for the uncertainty adaptation scheme, rather than for the role of the control methodology itself. From previous applications of time-delay control [13, 18, 19], it has been shown the time-delay approximation, which is the core idea of time-delay control, is a practical and efficient estimation method for unmodeled dynamics and uncertainties. In this paper, the trajectory optimization of the agile turn is performed in advance of autopilot design in order to discover the behavior of state variables and the practical considerations for autopilot design during the agile turn. These results can also be used for the reference trajectory during the agile turn phase in simulation studies. In order to investigate the performance of the proposed method, the proposed method is tested in 6-DOF nonlinear simulations with a step command and angle-of-attack profile obtained from the trajectory optimization. Furthermore, an intercept scenario, including the agile turn and the terminal homing phases, is also performed. For this purpose, the acceleration controller is also constructed based a cascaded control structure [20, 21]; the angle-of-attack controller is augmented by the outer proportional-integral (PI) controller; and at end of the agile turn phase, the autopilot controlling the angle-of-attack is reconfigured to control the body acceleration, using a simple control command blending logic.

The outline of this paper is as follows: Section II presents a nonlinear missile model considered in this work. In Section III, we analyze the agile turn maneuver in order to determine the behavior of state variables. We discuss the agile missile autopilot design, using the nonlinear backstepping control with time-delay adaptation methodology, in Section IV. In Section V, nonlinear simulations are performed to investigate the performance of the proposed method. Section VI summarizes this work.

## II. Missile Model

A nonlinear missile model with a boost [22] is considered, as shown in Fig. 1. It is assumed that the agile missile is maneuvering in the horizontal plane with the 90 degrees roll angle after its launch from an aircraft. Accordingly, the longitudinal gravity component is not considered. The longitudinal equation of motion in the body axis is written as

$$\dot{u} = -wq + (F_x + T)/m$$
$$\dot{w} = uq + F_z/m \quad (1)$$
$$\dot{q} = M_y/I_{yy}$$

where $u$, $w$, and $q$ represent the longitudinal body velocity, the vertical body velocity, and the body pitch angular velocity, respectively. $m$, $I_{yy}$, and $T$ denote mass, pitching moment of inertia, and thrust, respectively. The aerodynamic forces and moment, denoted by $F_x$, $F_z$, and $M_y$, are given as

$$F_x = -QSC_A, \qquad F_z = -QSC_N, \qquad M_y = QSlC_M \quad (2)$$

with

$$C_A \triangleq C_{A_0}(M) + C_{A_\alpha}(M)\alpha + C_{A_\delta}(M)(|\delta|/2)^2 + \Delta C_{A_T}(M)$$
$$C_N \triangleq C_{N_0}(\alpha, M) + C_{N_\delta}(\alpha, M)\delta + \Delta C_{N_\phi} \quad (3)$$
$$C_M \triangleq C_{M_0}(\alpha, M) + C_{M_q}(M)\frac{ql}{2V} + C_{M_\delta}(\alpha, M)\delta - C_N(x_{cg,ref} - x_{cg})/l + \Delta C_{M_\phi}$$

where $Q$, $S$, $l$, $M$, and $V$ are dynamic pressure, reference area, reference length, Mach number, and velocity, respectively. $\alpha$ and $\delta$ represent the angle-of-attack and the control fin deflection angle. $x_{cg,ref}$ and $x_{cg}$ denote the reference location of CG(center of gravity) at launch and the true location of CG during the booting phase. These parameters are given in Table. 1. In this missile model, $m$, $I_{yy}$, and $x_{cg}$ are linearly varied during the boosting phase due to the exhaustion of propellant.

### III. Autopilot Design

From Eq. (3), new variables are defined for convenience as follows:

$$x_1 = \alpha, \qquad x_2 = q, \qquad u = \delta \tag{4}$$

Substituting Eq. (3) into Eq.(4), we rewrite the dynamics in terms of state variables and control input separately:

$$\begin{aligned} \dot{x}_1 &= f_1 + x_2 + h_1(u) + g_1 \\ \dot{x}_2 &= f_2 + h_2 u + g_2 \end{aligned} \tag{5}$$

where

$$\begin{aligned} f_1 &\triangleq -\frac{QS}{mV}\left[C_{N_0}\cos\alpha - \left(C_{A_0} + C_{A_\alpha}\alpha + \Delta C_{A_T}\right)\sin\alpha\right] - \frac{T\sin\alpha}{mV}, \quad f_2 \triangleq \frac{QSl}{I_{yy}}\left[C_{M_0} + C_{M_q}\frac{ql}{2V} - C_{N_0}\frac{x_{cg,s}}{l}\right] \\ h_1(u) &\triangleq -\frac{QS}{mV}\left[C_{N_\delta}\delta\cos\alpha - C_{A_\delta}\left(|\delta|/2\right)^2\sin\alpha\right], \quad h_2 \triangleq \frac{QSl}{I_{yy}}\left[C_{M_\delta} - C_{N_\delta}\frac{x_{cg,s}}{l}\right] \\ g_1 &\triangleq -\frac{QS}{mV}\Delta C_{N_\phi}\cos\alpha, \quad g_2 \triangleq \frac{QSl}{I_{yy}}\Delta C_{M_\phi} \end{aligned} \tag{6}$$

and $x_{cg,s} \triangleq x_{cg,ref} - x_{cg}$. In the above equation, the contribution of $h_1(u)$ in the $\dot{x}_1$ equation can be ignored because of $|h_2 u| \gg |h_1(u)|$, generally, in the missile systems. The terms $g_1$ and $g_2$ are related with the aerodynamic roll coupling effect and reach their maximum values at $\phi = 45°$. In this study, the maximum values of these terms, a worst case scenario, are considered as the additive aerodynamics perturbations due to the induced roll effect in the high angle-of-attack. Also, taking the multiplicative aerodynamic uncertainties in Eq. (5) into consideration, the dynamic equation can be rewritten in a strict feedback structure with unknown terms as

$$\begin{aligned} \dot{x}_1 &= f_1 + x_2 + \Delta_1 \\ \dot{x}_2 &= f_2 + h_2 u + \Delta_2 \end{aligned} \tag{7}$$

where $\Delta_1$ and $\Delta_2$ can be regarded as the aerodynamic uncertainties caused by Eqs. (6) and (7), and the unmodeled dynamic related to $h_1(u)$. To construct the angle-of-attack controller based on the backstepping control methodology, let new residual variables be defined as

$$z_1 \triangleq x_1 - x_{1d}, \qquad z_2 \triangleq x_2 - x_{2d} \tag{8}$$

Taking the time-derivative of the residual variables yields

$$\dot{z}_1 = f_1 + x_2 + \Delta_1 - \dot{x}_{1d}$$
$$\dot{z}_2 = f_2 + h_2 u + \Delta_2 - \dot{x}_{2d} \qquad (9)$$

where $x_{1d}$ represents the desired values of $x_1$, which are determined by the reference command of the angle-of-attack. $x_{2d}$ is the desired value of $x_2$, and it forces $z_1$ to converge with zero in a finite time for backstepping control. Consider the following theorem:

Then, the control law for backstepping, which enforces $x_1$ and $x_2$, tracks their commands, $x_{1d}$ and $x_{2d}$, in a finite time, as can be determined by the following theorem.

From Theorem 1 and 2, it is noted that the control law and the desired command of $x_2$ contain the unknown terms $\Delta_1$ and $\Delta_2$, which causes the degradation of the tracking performance. In order to improve the tracking performance, $\Delta_1$ and $\Delta_2$ should be estimated and compensated for in regard to control commands as

$$x_{2d} = -f_1 - K_1 z_1 + \dot{x}_{1d} - \hat{\Delta}_1$$
$$u = h_2^{-1}\left(-f_2 - K_2 z_2 + \dot{x}_{2d} - z_1 - \hat{\Delta}_2\right) \qquad (10)$$

where $\hat{\Delta}_1$ and $\hat{\Delta}_2$ are the estimates of $\Delta_1$ and $\Delta_2$. Accordingly, we propose an efficient and practical adaptation scheme based on time-delay approximation [16]. The basic idea of time-delay approximation is that if a function $f(t)$ is continuous with regard to time for $0 \leq t \leq c$, then it allows the following for a sufficiently small time-delay $L$:

$$f(t) \cong f(t-L) \qquad (11)$$

In order to ensure the validity of this approximation, the unknown terms under consideration, $\Delta_1$ and $\Delta_2$, are also given by continuous functions of time. This is shown by the following lemma.

According to Lemma 1, the unknown terms $\Delta_1$ and $\Delta_2$ can be estimated using the time-delay approximation as

$$\hat{\Delta}_1 = \Delta_1(t-L), \qquad \hat{\Delta}_2 = \Delta_2(t-L) \qquad (12)$$

Using Eq. (7), the above equation can be written as

$$\hat{\Delta}_1 = \dot{x}_1(t-L) - f_1(t-L) - x_2(t-L)$$
$$\hat{\Delta}_2 = \dot{x}_2(t-L) - f_2(t-L) - h_2(t-L)u(t-L) \qquad (13)$$

As shown in Eq. (13), this adaptation scheme requires the time-delayed information regarding the state variables, control input, and models. The time-delayed information can be obtained from the single-lag system as

$$f(t-L) = \frac{1}{\tau_d s + 1} f(t) \quad (14)$$

where $\tau_d$ is the time constant value. The approximation error decreases as the time-delay constant value $\tau_d$ decreases. Since the estimation of $\Delta_1$ and $\Delta_2$ consists of continuous functions from Eq. (13), the control input in conjunction with $\hat{\Delta}_1$ and $\hat{\Delta}_2$ is also given by a continuous form, as shown in Eq. (10).

## IV. Simulation Results

In this section, the proposed autopilots for agile missiles are tested through three different 6-DOF nonlinear simulations. First, the basic characteristics and performance of the proposed autopilots are investigated by imposing step command. The second simulation analyzes the tracking performance of the proposed autopilots during the agile turn maneuvers. In this simulation, the angle-of-attack profile obtained from the trajectory optimization is used for the reference command of the angle-of-attack. Finally, the proposed autopilots are applied to engage a target in the rear hemisphere of the missile. In all simulations, it is assumed that the missile is launched at an altitude of 2000 m from an aircraft. The missile parameters shown in Table 1 are used. The second-order actuator models with natural frequency $\omega_{act} = 180\,\text{rad/s}$, damping ratio $\zeta_{act} = 0.7$, control fin limit $\delta_{\text{limit}} = \pm 30°$, and control fin rate limit $\dot{\delta}_{\text{limit}} = \pm 450°/\text{sec}$ are included. The design parameters for controllers are chosen as follows: $K_1 = K_2 = 25$, $\tau_d = 0.02$, $K_P = 0.0098$, and $K_I = 0.34$. In addition, the second-order linear command shaping filter is used for obtaining differential commands.

In this simulation, it is assumed that the missile starts on the agile turn with the initial velocity $V = 250\,\text{m/s}$ and is then boosted after a safe separation from an aircraft presents the step responses for angle-of-attack control, without considering the aerodynamic uncertainties. the solid lines and the dotted lines represent the response values and the command values, respectively. The results show the sound tracking performances of the proposed controller.

It gives the step responses for angle-of-attack control with 30% multiplicative uncertainties ($\Delta_{Pert} = 0.3$) in aerodynamic coefficients and additive aerodynamic uncertainties from the coupling effect in high angle-of-attack. the dotted line, the dashed line, and the solid line represent the angle-of-attack command, the angle-of-attack response without adaptation, and the angle-of-attack response with adaptation, respectively. The results indicate that even in the presence of aerodynamic uncertainties, the proposed controller can track the desired command quite well. The time-delay adaptation scheme can accurately estimate the unknown term.

## V. Conclusion

In this research, after investigating the agile turn maneuver based on the trajectory optimization, which revealed that the agile missile system can undergo highly nonlinear, rapidly changing dynamics and aerodynamic uncertainties during agile turns, we designed a nonlinear adaptive autopilot for controlling the angle-of-attack. The proposed controller was constructed based on the backstepping control methodology in conjunction with a practical time-delay adaptation scheme and tested in 6-DOF nonlinear simulations.